\documentstyle[aps,preprint]{revtex}
\begin{document}
\title { Coherence effects on pion spectrum distribution }
\author{Q.H. Zhang$^{1}$ \footnote{
        On leave from China Center of Advanced Science and Technology (CCAST),
P.O.Box 8730,Beijing 100080,P.R.China}, 
W.Q. Chao$^{2,3,4}$ and C.S. Gao$^{2,4,5}$}
\address{1 Institut f\"ur Theoretische Physik, Universit\"at Regensburg 
D-93040 Regensburg, Germany\\
        2 CCAST (World Laboratory) P.O. Box 8730, Beijing 100080,  China\\
	3 Institute of High Energy Physics, Academia Sinica, P.O. Box
	918(4),	Beijing 100039, China\\
	4 Institute of Theoretical Physics, Academia Sinica, P.O. Box 
	2735,	Beijing 100080, China\\
	5 Physics Department, Peking University,
         Beijing 100871, China}
\vfill
\maketitle

\begin{abstract}
The effects of two kinds of  coherent lengths,  the wave packet length 
of the emitter and 
the radius of the coherent source,  on pion spectrum distribution 
 are studied.  It is shown that both coherent lengths can 
 cause abundant pions at low momentum, but the DCC size effects on pion 
spectrum distribution is more important.  So observing abundant pions 
at low momentum may be taken as a signal of DCC effects.  
\end{abstract}

PACS number(s): 25.75+r, 13.85 Hd, 24.10-i.

		\section{Introduction}

Early in the next decade two heavy-ion accelerators, the 
Relativistic Heavy Ion Collider (RHIC) and the Large Hadron 
Collider (LHC), will create highly excited regions, similar 
to heavy-ion in size and with temperatures exceeding $200
MeV $.  Among the most interesting speculations regarding ultra-high 
energy heavy ion collisions there is the idea that regions of 
misaligned vacuum might occur\cite{Lee,RW,Gavin,AHW,AA,GGP}.  
In such misaligned regions, which are analogous to misaligned domains 
in a ferromagnet,  the chiral condensate points in a different 
direction from that favored in the ground state.  If they were 
produced, misaligned vacuum regions plausibly would behave as 
a pion laser, relaxing to the ground state by coherent 
pion emission\cite{Pratt,CGZ}.

It is generally assumed that the fluctuation of the ratio of neutral to 
charged pions may be viewed as a signature of DCC phenomena. If the DCC 
state is formed, the product of one kind of pions should be larger  
comparing to other kinds of pions.  Furthermore since a pion laser is formed, 
the mean momentum of the pions emitted from DCC regions should be 
much smaller.  PHOBOS\cite{BW} is a compact silicon detector designed to 
measure particle multiplicity for all charged particles and for 
photons and it will be used in AGS and RHIC. The abability
to measure photons will allow PHOBOS to study their parents $\pi^{0}$ and  
the fluctuation of the ratio of $\pi^{0}$ to 
$\pi^{+,-}$. The most interesting thing of PHOBOS for us is 
that it can measure the low mometum particles, that is  
it can detect the other signature of DCC phenomena.

Two-pion Bose-Einstein correlation is widely used in high energy 
heavy-ion collisions to provide the information of the space-time
structure, degree of coherence and dynamics of the region where 
the pions were produced\cite{BGJ,GKW,ZAJ,LOR}, which is closely  
related to the single and double pions inclusive distribution.
The coherent pion emission causes pions concentrated at low 
transverse momenta. This behavior should have explicit 
effects on the pion single particle spectrum and two-pion 
interferometry. Besides 
the large domain size of the disoriented chiral condensate
(DCC) regions, there is also another kind of 
coherent length which corresponds to the wave packet length 
scale of the emitter. The wave packet length should also affect the 
pion spectrum distribution. Then the question arises: which one,
the DCC size or the wavepacket length, affects more seriously 
on pion spectrum distribution.  To answer this 
question, in this paper, we first derive the 
formula of single particle spectrum by taking into account 
both DCC size and the wave packet length in Section two.  
As a simple example, 
the DCC region size and the emitter size 
effects on the pion spectrum distribution are given 
in Section three.   Conclusions are given in Section four.

\section{Pion spectrum distribution for partially coherent source}

It is widely accepted that a state created by a classical pion source 
is described by
\cite{BGJ,GKW,CGZ,ZQH1,ZQH}
\begin{equation}
|\phi>=exp(i\int d \vec{p} \int d^{4}x j(x) exp(ip\cdot x) c^{+}(\vec{p})
|0>,
\end{equation}
where $c^{+}(\vec {p})$ is the pion creation operator, $|0>$ is the pion 
vacuum.  $j(x)$ is the 
current of the pion, which can be expressed as 
\begin{equation}
j(x)=\int d^{4}x' d^{4}p j(x',p) \nu(x') exp(-ip\cdot (x-x')) .
\end{equation}
Here $j(x',p)$ is the probability amplitude of finding a pion 
with momentum $p$ , emitted by the emitter at $x'$. $\nu(x')$ 
is a random phase factor. All emitters are uncorrelated in 
coordinate space when assuming:
\begin{equation}
<\nu^{*}(x')\nu(x)>=\delta^{4}(x'-x)   .
\end{equation}
This is in ideal cases. In a more realistic case, each chaotic emitter has a
small coherent wave package length scale and the above 
equation can be replaced by 
\begin{equation}
<\nu^{*}(x')\nu(x)>=\frac{1}{\delta^{4}}
\exp\{-\frac{(x_{1}-x'_{1})^{2}}{\delta^{2}}
-\frac{(x_{2}-x'_{2})^{2}}{\delta^{2}}
-\frac{(x_{3}-x'_{3})^{2}}{\delta^{2}}
-\frac{(x_{0}-x'_{0})^{2}}{\delta^{2}} \}  .
\end{equation}
Here $\delta $ is a parameter which determines the coherent length (time) 
scale of the emitter.  For simplicity, the same coherent scale is taken for 
both spacial and time at the moment.  
The above formula means that two-emitters within the range of 
$\delta$ can be seen as one emitter, while two-emitters out of this
range are incoherent.  For simplicity we 
also assume that
\begin{equation}
<\nu^{*}(x)>=<\nu(x)>=0    ,
\end{equation}
which means that for each emitter the phases are randomly distributed in the 
range of $0$ to $2 \pi$. 

The coherent state can be expanded in Fock-Space as 
\begin{eqnarray}
|\phi>=\sum_{n=0}^{\infty}
\frac{(i \int  j(x) e^{ip\cdot x} c^{+}(p) d\vec{p} d^{4}x)^{n}}{n!}|0>
=\sum_{n=0}^{\infty}|n>,
\end{eqnarray}
with
\begin{equation}
|n>=\frac{(i \int  j(x) e^{ip\cdot x} c^{+}(p) d\vec{p} dx)^{n}}{n!}|0>.
\end{equation}

Here $|n>$ is the n-pion state, using the relationship
\begin{equation}
\left[c^+(\vec{p_1}), c(\vec{p_2})\right]=\delta(\vec{p_1}-\vec{p_2})
\end{equation}
we have 
\begin{equation}
c(\vec{p})|n>=i\int d^{4}x j(x) exp(ip\cdot x) |n-1>,
\end{equation}
then
\begin{eqnarray}
I(Q,K)&=&<1|c^{+}(\vec{p}_{1})c(\vec{p}_{2})|1>=
\int d^{4}x_{1} d^{4}x_{2} j^{*}(x_{1}) j(x_{2}) 
exp(-i(p_{1}\cdot x_{1}-p_{2}\cdot x_{2}))
\nonumber\\
&=&\int d^{4}x_1 d^{4}x_2 j^{*}(x_1)j(x_2)exp(-i(K/2+Q)\cdot x_1 -i(K/2-Q)\cdot x_2)
 \nonumber\\
&=&\int d^{4}x_1 d^{4}x_2 j^{*}(x_1)j(x_2)exp(-iK\cdot (x_1-x_2)/2 -iQ\cdot (x_1+x_2))
\\
&=&\int d^{4}y d^{4}Y j^{*}(Y+y/2)j(Y-y/2)exp(-iK\cdot y/2 -i2Q\cdot Y)
\nonumber\\
&=&\int d^{4}Y g_{w}(Y,k) exp(-i2Q\cdot Y)    .
\nonumber
\end{eqnarray}

Here $Y=\frac{x_{1}+x_{2}}{2}, y=x_{1}-x_{2}$ are four dimensional  
coordinates, while  
$ Q=\frac{p_{1}-p_{2}}{2}$ and 
 $K=2k=p_{1}+p_{2} $ are 
the corresponding four dimensional momenta.  The above transformation, 
 referred as Wigner transformation,   
can be found in the original paper of E. Wigner \cite{Wigner32}.  
$g_{w}(Y,K)$ is the Wigner 
function, which can be explained as the probability of finding a pion 
at $Y$ with momentum $k=K/2$\cite{Wigner32} and is defined as 
\begin{equation}
g_{w}(Y,k)=\int d^{4}y j^{*}(Y+y/2)j(Y-y/2)exp(-iK\cdot y/2) .
\end{equation}

Inserting eq.(2) into the above equation we have
\begin{eqnarray}
g_{w}(Y,k)&=&\int d^{4}y exp(-ik\cdot y)
\nonumber\\
	&&\int d^{4}x'j^{*}(x',p_{1})dp_{1}e^{ip_{1}\cdot (Y+y/2-x')}\nu^{*}(x')
\\
	&&\int d^{4}x''j(x'',p_{2})dp_{2}e^{-ip_{2}\cdot (Y-y/2-x'')}\nu(x'') ,
\nonumber
\end{eqnarray}
then the single pion inclusive distribution $P_{1}^{cha}(\vec{p})$ can be
expressed as (eq.10):
\begin{eqnarray}
P_{1}^{cha}(\vec{k})&=&<1|c^+(\vec{k})c(\vec{k})|1>=\int g_{w}(Y,k) d^{4}Y
\nonumber\\
	&=&\int d^{4}x' d^{4}x'' d^{4}p_{1}d^{4}p_{2} 
j^{*}(x',p_{1})j(x'',p_{2}) 
\nonumber\\
&&\nu^{*}(x')\nu(x'')\delta^{4}(k-\frac{p_{1}+p_{2}}{2})
\delta^{4}(p_{1}-p_{2})
e^{ip_{1}\cdot (Y-x')}e^{-ip_{2}\cdot (Y-x'')}\\
&=&\int d^{4}x' d^{4}x'' 
j^{*}(x',k)j(x'',k)\nu^{*}(x')\nu(x'')e^{-ik\cdot (x'-x'')}  .
\nonumber
\end{eqnarray}
with $k_{0}$ taken to be $k_{0}=\sqrt{\vec{k}^{2}+m_{\pi}^{2}}$.  In the above 
equation,  we have taken into account the wave packet length $\delta$ of the emitter 
which form a chaotic source.  

For a system with one finite size coherent 
source, e.g., a finite DCC region and many other totally chaotic 
emitters which form a 
chaotic source, the state is described by \cite{CGZ}
\begin{equation}
|\phi>_{part}=exp(i\int d \vec{p} \int d^{4}x (j(x)+j_{c}(x))
 exp(ip\cdot x) c^{+}(\vec{p})|0>~~~  ,
\end{equation}
where $c^{+}(p)$ is the pion creation operator; $ j_{c}(x)$ is the 
current of pions produced by the coherent source, which can be 
expressed as
\begin{equation}
j_{c}(x)=\int d^{4}x' d^{4}p j_{c}(x',p) \exp\{-ip\cdot (x-x')\};
\end{equation}
$j(x)$ is 
the current of pions produced by totally chaotic source , which can be 
expressed as eq.(2). The defference between $j(x)$ and $j_c(x)$ is 
that: For chaotic source, each emitter have different phase (different 
$\nu(x)$ in eq.(2)) while for $j_c(x)$ each emitter have the same phase.  
The state $|\phi>_{part}$ can be expanded as 
\begin{equation}
|\phi>=\sum_{n=0}^{\infty}
\frac{(i \int  (j(x)+j_{c}(x))e^{ip\cdot x} c^{+}(p) d\vec{p} dx)^{n}}{n!}|0>
=\sum_{n=0}^{\infty}|n>_{part}~~~~~,
\end{equation}
with
\begin{equation}
|n>_{part}=\frac{(i \int (j_{c}(x)+j(x)) e^{ip\cdot x} c^{+}(p) 
d\vec{p} d^4x)^{n}}{n!}|0>.
\end{equation}
Here $|n>_{part}$ is the n-pion state.  Then the pion's spectrum distribution 
is 
\begin{eqnarray}
P_{1}^{part}(\vec{p})&=&_{part}<1|c^{+}(\vec{p})c(\vec{p})|1>_{part}
\nonumber\\
&=&\left( \int d^4 x (j(x) +j_c(x))e^{ip\cdot x} \right)^*
\left( \int d^4 x (j(x) +j_c(x))e^{ip\cdot x} \right)
\\
&=&\int d^4x_1d^4x_2 (j^*(x_1)j(x_2)+j^*(x_1)j_c(x_2)
+j^*_c(x_1)j(x_2)+
\nonumber\\
&&j^{*}_{c}(x_1)j_c(x_2) ) e^{-ip\cdot (x_1-x_2)}
\nonumber
\end{eqnarray}
Taking the phase average and using the relationship of eq.(5) 
we have
\begin{equation}
<j^*(x)j_c(y)>=<j_c(x)j^*(y)>=0
\end{equation}
Then the above equation can be re-expressed as
\begin{eqnarray}
P_1(\vec{p})&=&\int d^4x_1d^4 x_2 j^*(x_1)j(x_2) e^{-i(p\cdot (x_1 - x_2))}
\nonumber\\
&&	+\int d^4x_1d^4x_2 j^*_c(x_1)j_c(x_2)e^{-i(p\cdot (x_1- x_2))}
\\
&=&\int g_w(x,\vec{p})d^4 x + |j_c(\vec{p})|^2
\nonumber
\end{eqnarray}
Here $j_{c}(\vec{p})$ can be expressed as
\begin{equation}
j_{c}(\vec{p})=\int j_{c}(x) \exp(ip\cdot x) d^{4} x   .
\end{equation}
The above formula (eq.(20)) shows that the total pion spectrum 
distribution is consist of two-parts, one is spectrum distribution 
of the chaotic source, the other is 
the spectrum distribution of the coherent source.  
In the above derivation we have taken into account 
the wave packet length and the coherent pion source $j_{c}$, 
therefore, in our formulation it is possible to examine both the wave 
packet length of each chaotic emitter and the DCC radius effects 
on the pion single particle distribution when the pions emitted from the 
DCC region are assumed to be coherent. 
 
\section{Coherent length effects on pion spectrum }

In this section, we will give an example to investigate the 
wave packet length and the coherence source radius effects on 
single pion distributions.  
We assume that the chaotic emitter amplitude distribution is
\begin{equation}
j(x,k)=\exp(\frac{-x_{1}^{2}-x_{2}^{2}-x_{3}^{2}}{2R_{0}^{2}})
\delta(x_{0})
exp(-\frac{k_{1}^{2}+k_{2}^{2}+k_{3}^{2}}{2\Delta^{2}})~~~.
\end{equation}
Where $R_{0}$ and $  \Delta $ are parameters which 
represent the radius of the chaotic source size and the momentum 
range of pions respectively. Here $x=(x_0,x_1.x_2,x_3)$ 
and $k=(k_0,k_1,k_2,k_3)$ 
is pion's coordinate and momentum respectively. 
Bringing  eq.(4) and eq.(22) into 
eq.(13), we can easily get the pion single 
particle spectrum distribution
\begin{equation}
P_{1}^{cha}(\vec{p})=
(\frac{\frac{1}{\Delta^{2}}+\frac{R_{0}^{2}\delta^{2}}{\delta^{2}+4R_{0}^{2}}
}{\pi})^{\frac{3}{2}}
\exp\{-\vec{p}^{2}\cdot 
(\frac{1}{\Delta^{2}}+\frac{R_{0}^{2}\delta^{2}}{\delta^{2}+4R_{0}^{2}}
)\}  .
\end{equation}

From the above expressions, we can see that the wave packet length of 
each chaotic emitter has 
great influence on the pion single particle inclusive distribution.  
The single particle momentum distribution is shown in fig.1 .  
The input value of $R_{0}$ and $\Delta$ is 
$5 fm $ and $0.3 GeV $, respectively.  
The solid, dashed and dot-dashed lines correspond to 
$\delta =0 fm$,  $0.5 fm $ and
$ 1 fm $ , respectively. 
It is clearly shown that as the wave packet length $\delta$ increases, 
the mean momentum of pions gets smaller, which means that the wave packet 
length of the chaotic emitter can cause abundant pions at low momentum.   

Now we consider the finite size coherence source  effects on the 
pion single 
particle inclusive distribution.  We assume that 
the emitting amplitude of coherent pions is 
\begin{equation}
j_{c}(x,k)=\exp(\frac{-x_{1}^{2}-x_{2}^{2}-x_{3}^{2}}{2R_{c}^{2}})
\delta(x_{0})
exp(-\frac{k_{1}^{2}+k_{2}^{2}+k_{3}^{2}}{2\Delta_{c}^{2}}) .
\end{equation}
Here $R_{c}$and $  \Delta_{c} $ are parameters which 
represent the radius of the coherent source, e.g. the 
DCC region, and the mean momentum of coherent pions. 
$x=(x_0,x_1,x_2,x_3)$ and $k=(k_0,k_1,k_2,k_3)$ is pion's 
coordinate and momentum respectively.  
The normalized pion single particle distribution can be 
expressed as 
\begin{equation}
P_{nor}^{part}(\vec{p})=
A\cdot P^{cha}_{nor}(\vec{p}) + (1-A) P^{c}_{nor}(\vec{p})~~~~~,
\end{equation}
where $ A $ is a parameter which determines the incoherent degree of the 
source and is defined by
\begin{equation}
A=\frac{\int d\vec{p} P^{cha}_{1}(\vec{p})} {\int d\vec{p} 
(P^{cha}_{1}(\vec{p}) + P^{c}_{1}(\vec{p}))}.  
\end{equation}
For $A=1$ the source is totally chaotic,  for $A=0$ the source is 
totally coherent, otherwise the source is partially coherent. 
Here $P^{cha}_{nor}(\vec{p})$ can be expressed as
\begin{equation}
P^{cha}_{nor}(\vec{p})=
(\frac
{\frac{1}{\Delta^{2}}+\frac{R_{0}^{2}\delta^{2}}{\delta^{2}+4R_{0}^{2}}
}{\pi})^{\frac{3}{2}}
\exp\{-\vec{p}^{2}(\frac{1}{\Delta^{2}}+
\frac{R_{0}^{2}\delta^{2}}{\delta^{2}+4R_{0}^{2}}
)\}
\end{equation}
and $P^{c}_{nor}(\vec{p})$ can be expressed as 
\begin{equation}
P^{c}_{nor}(\vec{p})=(\frac
{\frac{1}{\Delta_{c}^{2}}+R_{c}^{2}}
{\pi})^{3/2}
\exp\{-\vec{p}^{2}(\frac{1}{\Delta_{c}^{2}}+R_{c}^{2}) \}~~~.
\end{equation}
Then the single particle inclusive distribution for partially coherent 
source is shown in fig.2,  
where the input values of $A$,  $R_{0}$,  $\Delta$,  $\Delta_{c}$ and 
$\delta$  is
$0.5$,  $5 fm$,  $0.3 GeV$,  $0.15 GeV $ and $0.3 fm$,  respectively.  
The solid, dashed and dot-dashed lines correspond to 
$R_{c}=1fm$,  $2fm$ and $ 3fm$, respectively.
It is clear that as $R_{c}$ becomes larger, that is 
DCC region becomes larger, the pion's mean momentum becomes 
smaller, this condition is consistent with the nature of the 
coherent property of the source.  It can be seen from fig.2 that 
DCC effects on pion spectrum distribution is more important 
than emitter size effects. So observing abundant pions at low 
momentum can be taken as a signature of DCC effect.

\section {Conclusions}

It has been suggested that a large DCC region may be formed in 
relativistic heavy-ion collisions.   If the DCC region is formed, a 
large number of lower momentum pions should be produced, which is 
one of the signature of DCC phenomena. 
There is also another kind of coherent length 
which corresponds to the size of the wave packet length of each 
chaotic emitter and which also affects the pion momentum 
distribution.  Therefore it is very interesting to analyze 
the effects of the two kind of lengths, namely the
DCC size and wave packet length, on pion single particle inclusive 
distributions and to find out whose effect is more imporatnt. 

In this paper, as a simple example,  we have derived the formula 
of the pion spectrum distribution  by taking into account both 
the wavepacket length and DCC size
 and analyzed the effect of the two coherent lengths on 
pion inclusive distributions.   
It has been shown that both coherent lengths can cause 
 the  abundance of  pions at low momentum.  Among the two, 
the DCC size effects on the pion spectrum distribution is more important.
  Therefore observing 
abundant pions at low momentum may provide a signal 
of the DCC effects.  Such a signal can be detected by PHOBOS at RHIC.

\section*{{Acknowledgement}}

The authors would like to express their gratitude to the referees 
for their helpful suggestion and comments. One of the authors (Q.H.Z.)
would like to express his thanks to  
Dr. Pang Yang for helpful discussions. 
This study was partially supported by the National Natural Science 
Foundation of China, Post-doctoral Science Foundation of China and 
the Alexander von Humboldt Foundation in Germany .

\newpage

\section*{Figure Captions}
\parindent=0pt

Fig.1 The pion inclusive distribution. The input values of $R_{0}$ and
      $\Delta$ are $ 5 fm $ and $0.3 GeV$. The solid, dashed and 
      dot-dashed lines corresponds to $\delta = 0 fm$ ,  $0.5 fm$ and
      $1 fm $,  respectively.  

Fig.2 The pion inclusive distribution for partial coherent source. 
      The input values of $R_{0}$,  $\Delta$,  $\Delta_{c}$,  $\delta$ 
      and $A$ are 
      $5 fm$,  $0.3 GeV$,  $0.15 GeV$,  $0.3 fm $ and $0.5$ .  
      The solid, dashed and dot-dashed 
      lines corresponds to $R_{c}=1 fm$,  $ 2 fm$ and $3 fm $, respectively.

\end{document}